\documentclass[10pt,aps,english]{revtex4}
\usepackage[T1]{fontenc}
\usepackage[latin9]{inputenc}
\usepackage{amsmath}
\usepackage{amssymb}
\usepackage{graphicx}
\usepackage{babel}
\usepackage{pifont}

\makeatletter
\@ifundefined{textcolor}{}
{
 \definecolor{BLACK}{gray}{0}
 \definecolor{WHITE}{gray}{1}
 \definecolor{RED}{rgb}{1,0,0}
 \definecolor{GREEN}{rgb}{0,1,0}
 \definecolor{BLUE}{rgb}{0,0,1}
 \definecolor{CYAN}{cmyk}{1,0,0,0}
 \definecolor{MAGENTA}{cmyk}{0,1,0,0}
 \definecolor{YELLOW}{cmyk}{0,0,1,0}
 }
\makeatother

\begin{document}

\title{Direct and Parallel Tomography of a Quantum Process}
\author{Yu-Xiang Zhang$^1$, Shengjun Wu$^{1,2}$, and Zeng-Bing Chen$^1$}

\affiliation{$^1$Hefei National Laboratory for Physical Sciences at Microscale, Department
of Modern Physics, and the Collaborative Innovation Center for Quantum
Information and Quantum Frontiers, University of Science and Technology
of China, Hefei, Anhui 230026, China. \\
$^2$Kuang Yaming Honors School, Nanjing Univeresity, Nanjing, Jiangsu 210093, China.}

\begin{abstract}
As the method to completely characterize quantum dynamical processes,
quantum process tomography (QPT) is vitally important for quantum
information processing and quantum control, where the faithfulness
of quantum devices plays an essential role. Here via weak measurements,
we present a new QPT scheme characterized by its directness and parallelism.
Comparing with the existing schemes, our scheme needs a simpler state
preparation and much fewer experimental setups. Furthermore, each
parameter of the quantum process is directly determined from only
five experimental values in our scheme, meaning that our scheme is
robust against the accumulation of errors.
\end{abstract}
\maketitle

High-precision operations in quantum information processing and quantum
control vitally depend on the faithfulness of quantum devices. How
to completely characterize quantum dynamic evolutions within the quantum
devices is the task of quantum process tomography (QPT), which is
a topic of essential importance. Many theoretical and experimental
works on QPT have been done \cite{AAPT,AAPT 2,DCQD,experiment tomography 1,QPT experiment,QPT standard,science optics,science symmetry characterization,SEQPT,standard 2,DCQD2}.

A quantum process (QP) $\mathcal{E}$ can be viewed as a linear super-operator
that maps a state $\rho$ on the input Hilbert space $\mathcal{H}_{in}$
(of dimension $d_{in}$) to the state $\mathcal{E}(\rho)$ on the
output Hilbert space $\mathcal{H}_{out}$ (of dimension $d_{out}$).
To express it explicitly, we choose two sets of basis states $\{\left\vert \psi_{i_{1}}\right\rangle |i_{1}=1,\cdots,d_{in}\}$
and $\{\left\vert \alpha_{i_{2}}\right\rangle |i_{2}=1,\cdots,d_{in}\}$
in $\mathcal{H}_{in}$, and two sets of basis states $\{\left\vert \beta_{i_{3}}\right\rangle |i_{3}=1,\cdots,d_{out}\}$
and $\{\left\vert \phi_{i_{4}}\right\rangle |i_{4}=1,\cdots,d_{out}\}$
in $\mathcal{H}_{out}$. A QP $\mathcal{E}$ (more generally, any
linear super-operator) is associated with an operator $\mathrm{E}=\sum_{i_{1},i_{2}=1}^{d_{in}}\sum_{i_{3},i_{4}=1}^{d_{out}}\chi_{i_{1}i_{2}i_{3}i_{4}}\left\vert \psi_{i_{1}}\right\rangle \left\langle \alpha_{i_{2}}\right\vert \otimes\left\vert \beta_{i_{3}}\right\rangle \left\langle \phi_{i_{4}}\right\vert $,
and it maps an arbitrary operator $\Omega_{in}$ on $\mathcal{H}_{in}$
to an operator on $\mathcal{H}_{out}$: $\Omega_{in}\rightarrow\mathcal{E}(\Omega_{in})=\mathrm{tr}_{in}[(\Omega_{in}\otimes I)\mathrm{E}]$,
with the trace taken over the input Hilbert space only, i.e.,
\begin{equation}
\mathcal{E}(\Omega_{in})=\sum_{i_{1},i_{2}=1}^{d_{in}}\sum_{i_{3},i_{4}=1}^{d_{out}}\chi_{i_{1}i_{2}i_{3}i_{4}}\left\langle \alpha_{i_{2}}\right\vert \Omega_{in}\left\vert \psi_{i_{1}}\right\rangle \left\vert \beta_{i_{3}}\right\rangle \left\langle \phi_{i_{4}}\right\vert .\label{processdefinition}
\end{equation}
When the four sets of basis states are chosen, the quantum process
$\mathcal{E}$ is completely defined by the $d_{in}^{2}d_{out}^{2}$
complex parameters $\chi_{i_{1}i_{2}i_{3}i_{4}}$, which need to satisfy
certain conditions to ensure that the linear map defined in (\ref{processdefinition})
is a completely positive map and $\mathcal{E}(\rho)$ is really a
density matrix.

Tomography of an unknown QP $\mathcal{E}$ is, therefore, to determine
all the coefficients $\chi_{i_{1}i_{2}i_{3}i_{4}}$ with respect to
the chosen bases, such that one can predict the output state for any
input state. To accomplish it, the existing schemes \cite{AAPT,AAPT 2,DCQD,DCQD2,SEQPT,standard 2,QPT standard}
generally require measurements of many non-commuting observables and
a large number of different input states, and thus become impractical
when the dimension of the quantum system increases.

A QPT scheme provides a way to express the QP parameters in terms
of expectation values of some observables, which can be directly obtained
from experiments. So we consider a QPT scheme {}``more direct\textquotedblright{}
if it requires fewer expectation values to determine a single QP parameter.
We seek a QPT scheme that establishes a direct connection between
the QP parameters and the experimental expectation values. Such a
direct scheme is efficient when applied to partial QPT: to obtain
a single or a few parameters of the QP, one needs to perform only
the relevant measurements instead of all the measurements required
for a complete tomography of all the parameters \cite{selective}.
Another advantage is that a direct scheme is also robust against the
accumulation of errors as it relates a QP parameter to fewer experimental
expectation values.

Generally, a QPT scheme requires various input states, and on each
input state some non-commuting observables should be measured. These
non-compatible measurements cannot be performed simultaneously, i.e.,
cannot be performed in a single experimental setup. From one setup
to another, the experimenters have to change and recalibrate the devices.
Hence for a QPT scheme, the number of setups closely relates to the
efficiency of a complete tomography. For example, in the standard
QPT scheme \cite{QPT standard,standard 2} one setup gives only one
expectation value, and as such, even for the tomography of a{}{}``simple\textquotedblright{}
two-qubit gate, it requires about $4^{4}=256$ different setups. To
improve the efficiency, it is desirable to make the measurements compatible,
so that much more expectation values can be obtained simultaneously
in a single setup. In another word, the expectation values required
to determine QP parameters should be obtained in parallel, not in
sequence. Such a feature is referred to as {}``parallelism\textquotedblright{}.
More parallel a QPT scheme is, fewer setups it requires. Clearly,
as a quantification of parallelism, the number of setups required
in a QPT scheme also depends on the number of different input states,
since different states need different methods to prepare.

In this letter, we shall present a direct and parallel QPT scheme
via weak measurements, which has already been demonstrated and used
for various purposes theoretically and experimentally \cite{first weak experiment,key,key-1,key-3,quantum dots,weak experiment,weak measurement AAV,science onepage,pnas leggett inequality,naphys bell inequality},
including the characterization of quantum states \cite{also state tomography,wave function,state tomography,state}.
Our scheme is direct since a single QP parameter is determined by
only five expectation values, regardless of the dimension of the quantum
system. It is parallel because the scheme requires only $d_{in}$
experimental setups. Our scheme also has a simple input state preparation,
we need only $d_{in}$ different input states, and for the tomography
of multi-particle processes we just need product input states.

\begin{figure}[tbp]
 \includegraphics[width=8.7cm]{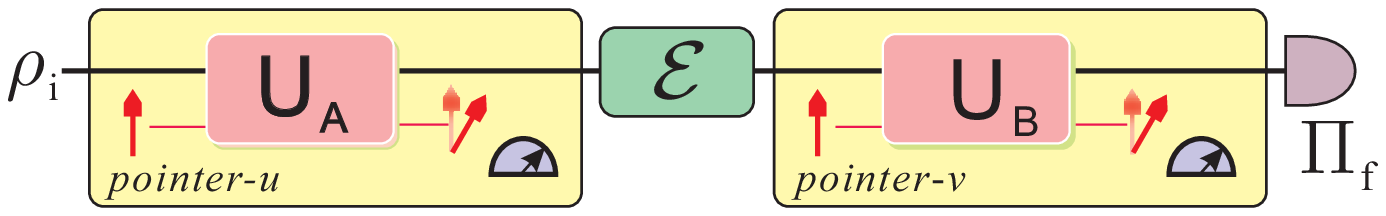} \caption{The main scheme. The system is initially prepared in a state $\rho_{i}$
(usually chosen from an orthogonal basis $\{\left\vert \psi_{i}\right\rangle \}$),
then weakly interacts with a pointer {}{}``$u$\textquotedblright{},
undergoes the unknown quantum process $\mathcal{E}$, weakly interacts
with another pointer {}{}``$v$\textquotedblright{}, and then is
post-selected by a projective measurement $\{\Pi_{f}\}$. The two
weak measurements associate with observables $\hat{A}$ and $\hat{B}$,
respectively. The pointers can be either continuous-variable systems
or discrete-variable systems. Joint read-out of the two pointers is
required to obtain the $X$-value, with which one can obtain the corresponding
QP parameter. The directness of our scheme is reflected in the result
that one QP parameter associates with one specific combination of
$\rho_{i}$, $\hat{A}$, $\hat{B}$ and $\Pi_{f}$.}
\end{figure}

The key features of our scheme are depicted in Fig. 1. Firstly, we
choose a system of dimension $d_{in}$ and prepare it in a state $\rho_{i}$,
then weakly couple it with a pointer labeled by the letter{}{}``$u$\textquotedblright{}.
The interaction between them is described by the Hamiltonian $H_{A}=g\delta(t-t_{0})\hat{A}\otimes\hat{p}_{u}$,
which results in a unitary evolution $U_{A}=\exp(-ig\hat{A}\otimes\hat{p}_{u})$.
A small $g$ is chosen to ensure that the interaction is weak enough.
Next, the system undergoes the unknown QP $\mathcal{E}$. Then, the
system weakly interacts with a second pointer (labeled by {}``$v$\textquotedblright{}),
the evolution is described by $U_{B}=\exp(-i\lambda\hat{B}\otimes\hat{p}_{v})$
where a small $\lambda$ is chosen to ensure a weakened interaction.
Finally, a projection $\Pi_{f}$ (the{}{}``post-selection\textquotedblright{})
is performed on the system. We use the notation $(i\hat{A}\hat{B}f)$
to denote this particular set of actions on the system.

For simplicity, suppose that the pointers {}``$u$\textquotedblright{}\ and{}{}``$v$\textquotedblright{}\ are
initially prepared in the same state $\sigma$. We also suppose that
the average values of two non-commuting observables $\hat{q}$ and
$\hat{p}$ of each pointer are initially zero, i.e., $\mathrm{tr}(\hat{p}\sigma)=\mathrm{tr}(\hat{q}\sigma)=0$.
For each pointer, we can read out either its $\hat{q}$ or $\hat{p}$.
Therefore, there are four possible products of pointer shifts, and
we define their averages as: $r_{1}=\langle\hat{p}_{u}\otimes\hat{p}_{v}\rangle$,
$r_{2}=\langle\hat{p}_{u}\otimes\hat{q}_{v}\rangle$, $r_{3}=\langle\hat{q}_{u}\otimes\hat{p}_{v}\rangle$
and $r_{4}=\langle\hat{q}_{u}\otimes\hat{q}_{v}\rangle$ for each
set $(i\hat{A}\hat{B}f)$ of actions on the system. These $r$-values
can be directly obtained from the experiments.

For this unknown process $\mathcal{E}$, the initial state $\rho_{i}$
of the system and the final post-selection $\Pi_{f}$, we can define
an{}{}``$X$-value\textquotedblright{}\ of $\hat{A}$ and $\hat{B}$
as

\begin{equation}
\begin{aligned}X^{i\hat{A}\hat{B}f}=\mathrm{tr}[\Pi_{f}\hat{B}\mathcal{E}(\hat{A}\rho_{i})]\end{aligned}
\label{eq:define X-1}
\end{equation}
 In the Supplementary Information, we show that when the set of actions
for the system is $(i\hat{A}\hat{B}f)$, the $X$-value $X^{i\hat{A}\hat{B}f}$
is related to the corresponding four $r$-values, via
\begin{equation}
X^{i\hat{A}\hat{B}f}=\frac{p_{f|i\hat{A}\hat{B}}}{4(\mathrm{Im}c_{1})^{2}g\lambda}[\mbox{ }(\frac{c_{1}^{\ast}}{c_{2}})^{2}r_{1}-\frac{c_{1}^{\ast}}{c_{2}}(r_{2}+r_{3})+r_{4}\mbox{ }]\label{x-r-relation}
\end{equation}
 where $p_{f|i\hat{A}\hat{B}}$ denotes the probability of the post-selection
$\Pi_{f}$, which can be obtained by statistics, and both $c_{1}=\mathrm{tr}(\hat{q}\hat{p}\sigma)$
and $c_{2}=\mathrm{tr}(\hat{p}\hat{p}\sigma)$ depend only on the
initial state of the pointers. For example, if we use continuous pointers
that are initially prepared in a Gaussian wave package $\frac{1}{(2\pi\Delta^{2})^{1/4}}\exp(-\frac{q^{2}}{4\Delta^{2}})$
with $[\hat{q},\hat{p}]=i$, we have $c_{1}=\frac{1}{2}i$ and $c_{2}=\frac{1}{4\Delta^{2}}$.
If we use qubit pointers that are initially prepared in $|\uparrow\rangle$
(the eigenstate of $\hat{\sigma}_{z}$ with eigenvalue $+1$), with
the replacement $\hat{p}\rightarrow\hat{\sigma}_{x}$ and $\hat{q}\rightarrow\hat{\sigma}_{y}$,
then we have $c_{1}=-i$ and $c_{2}=1$.

In order to determine the parameter $\chi_{i_{1}i_{2}i_{3}i_{4}}$
of the QP $\mathcal{E}$ in (\ref{processdefinition}), we prepare
the system in the $i_{1}$-th basis state $\left\vert \psi_{i_{1}}\right\rangle $,
and choose the first observable of weak measurement as $\hat{A}=\left\vert \alpha_{i_{2}}\right\rangle \left\langle \alpha_{i_{2}}\right\vert $,
and the second observable as $\hat{B}=\left\vert \beta_{i_{3}}\right\rangle \left\langle \beta_{i_{3}}\right\vert $.
Conditional on the final post-selected state $\left\vert \phi_{i_{4}}\right\rangle $,
we average the four possible products of pointer shifts to obtain
the $r$-values, from which the $X$-value $X^{i_{1}i_{2}i_{3}i_{4}}$
can also be obtained easily via (\ref{x-r-relation}). In Supplementary
Information, we show that
\begin{equation}
\chi_{i_{1}i_{2}i_{3}i_{4}}=\frac{1}{\langle\phi_{i_{4}}|\beta_{i_{3}}\rangle\langle\alpha_{i_{2}}|\psi_{i_{1}}\rangle}X^{i_{1}i_{2}i_{3}i_{4}}.\label{chi-X-1-relation}
\end{equation}
 The denominator in equation (\ref{chi-X-1-relation}) is fixed when
the bases are fixed.

To determine all the QP parameters, our scheme requires $d_{in}$
different input states $\{\left\vert \psi_{i_{1}}\right\rangle \}$
which compose an orthonormal basis in $\mathcal{H}_{in}$. Furthermore,
when the QP is a multi-particle one, we only need to prepare the multi-particle
system in product states, because we can choose a set of product basis
states to write the QP parameters as in equation (\ref{processdefinition}).
The final post-selection in our scheme is a complete projective measurement
onto the basis $\{\left\vert \phi_{i_{4}}\right\rangle \}$. In contrast
to many other applications of weak measurements, we do not throw away
any data via post-selection, and all data are used for tomography.

We also need weak measurements of $d_{in}$ different observables
$\{\left\vert \alpha_{i_{2}}\right\rangle \left\langle \alpha_{i_{2}}\right\vert |i_{2}=1,\cdots,d_{in}\}$
on the input system, and weak measurements of $d_{out}$ different
observables $\{\left\vert \beta_{i_{3}}\right\rangle \left\langle \beta_{i_{3}}\right\vert |i_{3}=1,\cdots,d_{out}\}$
on the system output from the unknown QP. All of these observables
can be measured simultaneously in one setup as the change of system
state due to each weak measurement is negligible. Suppose that coupling
constants of all the weak measurements have the same order of $\epsilon$,
then the principal contributions to an $r$-value of two pointers
that associate with observables $\hat{A}$ and $\hat{B}$ are of order
$\epsilon^{2}$, while the influence to it caused by an additional
pointer is of order $\epsilon^{4}$. This result also ensures that
we can introduce two pointers to weakly measure a same observable,
and finally read out the $\hat{q}$ shift of one pointer and the $\hat{p}$
shift of the other pointer. These features are shown in Fig. 2. In
such a way, the $4d_{in}d_{out}^{2}$ $r$-values that correspond
to one specific input state, the $d_{in}$ observables on $\mathcal{H}_{in}$,
the $d_{out}$ observables on $\mathcal{H}_{out}$ and the $d_{out}$
post-selected states, can be obtained in a single experimental setup.
In other words, $d_{in}d_{out}^{2}$ QP parameters can be determined
simultaneously. So our QPT scheme is a parallel one, the number of
setups equals the number of different input states, which is $d_{in}$.

\begin{figure}[tbp]
 \includegraphics[clip,width=8.7cm]{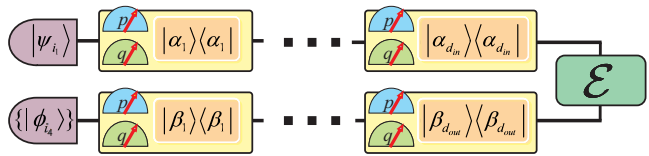} \caption{Parallelism. All the weak measurements can be performed simultaneously
in a single setup, i.e., in parallel. The $\hat{p}$ and $\hat{q}$
of a pointer can be read out simultaneously by introducing a duplicate
pointer. }
\end{figure}

In our scheme, the QP parameters are directly related to the experimental
data in an elegant way, as revealed in equations (\ref{x-r-relation})
and (\ref{chi-X-1-relation}): the parameter $\chi_{i_{1}i_{2}i_{3}i_{4}}$
is completely determined by the $X$-value $X^{i_{1}i_{2}i_{3}i_{4}}$,
which, in turn, is determined by five experimental values, i.e., the
four $r$-values and the probability $p_{f|i\hat{A}\hat{B}}$. This
result is valid regardless of the dimension of the system. Immediately,
it leads to the fact that our scheme is very efficient for partial
QPT, in which not all QP parameters are of interest. Furthermore,
the direct relation established by our scheme could sharply reduce
the accumulation of experimental systematic errors, which may lead
to a non-physical reconstruction of the unknown QP \cite{self consistent QPT,non-positive}.
We use $\delta$ to roughly denote each systematical error produced
in the experimental operations, including the state preparation and
measurements. Then for our QPT scheme, it can be shown from equation
(\ref{eq:define X-1}) and equation (\ref{chi-X-1-relation}) that
the accumulated systematic error is given as $\delta\chi_{i_{1}i_{2}i_{3}i_{4}}\sim\delta(|\psi_{i_{1}}\rangle)+\delta(|\alpha_{i_{2}}\rangle)+\delta(|\beta_{i_{3}}\rangle)+\delta(|\phi_{i_{4}}\rangle)$.
The number of error terms is a constant, which does not depend on
the dimensions of $\mathcal{H}_{in}$ and $\mathcal{H}_{out}$.

As to the existing QPT schemes \cite{AAPT,AAPT 2,DCQD,DCQD2,QPT standard,SEQPT,standard 2},
we present in details in the Supplemental Material that, the numbers
of required setups are generally of order $O(d^{2})$, $O(d^{4})$
or $O(d^{6})$ \cite{review A,DCQD2,SEQPT}, where we have assumed
that $d_{in}=d_{out}=d$. The accumulated error of one QP parameter
is generally $\delta(\chi)\sim2d^{2}\delta$ or $\sim2d^{4}\delta$.
The number of different input states required by them is generally
$O(d^{2})$ \cite{standard 2,SEQPT,QPT standard,DCQD}. Although in
some schemes the number of input states is of the same order with
ours \cite{DCQD2,AAPT,AAPT 2}, an ancilla system has to be introduced,
and the initial states must be correlated states of the combined system,
which are much more difficult to prepare. In the Supplemental Material
we also propose an ancilla-assisted version of our scheme, it requires
only one input state and one setup. Based on these facts, the advantages
of our schemes are apparent.

We also present further extensions of our scheme in the Supplemental
Material. When the coupling constants such as $\lambda$ and $g$
are not so small, i.e., the measurements are not so weak, a tomography
scheme of the QP is also presented, and the problem can actually be
solved exactly since all the observables we have considered so far
are projectors. One can also consider observables other than projectors,
we illustrate this point via an example of the tomography of a single-qubit
gate in the Supplemental Material, where we use one Pauli matrix as
the observable and one observable is sufficient. An extension of our
scheme for the case of multi-particle QPT with only weak measurements
performed on single particles is also proposed. In the scheme, only
product input states are required.

In summary, our scheme requires the simplest state preparation, it
is more parallel than the existing schemes and requires the fewest
setups. Being the most direct scheme, our scheme is also robust against
error accumulation. Weak measurements, the building blocks in our
scheme, have already been demonstrated and used for various purposes
experimentally \cite{first weak experiment,key,key-1,key-3,weak experiment,wave function,naphys bell inequality,pnas leggett inequality,science onepage}.
So our scheme can be implemented under current technology.

This work was supported by the National Natural Science Foundation
of China (Grants Nos. 11275181 and 61125502), the National Fundamental
Research Program of China (Grant No. 2011CB921300), the Chinese Academy
of Sciences, and the National High Technology Research and Development
Program of China.

\section{supplemental material}
As stated in the main text, we like to have a tomography scheme for
an unknown quantum process (QP) $\mathcal{E}$, which can be viewed
as a linear super-operator that maps an input state $\rho_{in}$ on
$\mathcal{H}_{in}$ (of dimension $d_{in}$) to the output state $\mathcal{E}(\rho)$
on $\mathcal{H}_{out}$ (of dimension $d_{out}$). Here, $d_{in}\neq d_{out}$
in general.

In order to express the QP explicitly, we choose two sets of basis
states $\{\left|\psi_{i_{1}}\right\rangle |i_{1}=1,\cdots,d_{in}\}$
and $\{\left|\alpha_{i_{2}}\right\rangle |i_{2}=1,\cdots,d_{in}\}$
in the input Hilbert space $\mathcal{H}_{in}$, and two sets of basis
states $\{\left|\beta_{i_{3}}\right\rangle |i_{3}=1,\cdots,d_{out}\}$
and $\{\left|\phi_{i_{4}}\right\rangle |i_{4}=1,\cdots,d_{out}\}$
in the output Hilbert space $\mathcal{H}_{out}$. Any QP $\mathcal{E}$
(more generally, any linear super-operator) is associated with an
operator ${\rm E}=\sum_{i_{1},i_{2}=1}^{d_{in}}\sum_{i_{3},i_{4}=1}^{d_{out}}\chi_{i_{1}i_{2}i_{3}i_{4}}\left|\psi_{i_{1}}\right\rangle \left\langle \alpha_{i_{2}}\right|\otimes\left|\beta_{i_{3}}\right\rangle \left\langle \phi_{i_{4}}\right|$,
and maps an arbitrary operator $\Omega_{in}$ on $\mathcal{H}_{in}$
to an operator on $\mathcal{H}_{out}$: $\Omega_{in}\rightarrow\mathcal{E}(\Omega_{in})=tr_{in}[(\Omega_{in}\otimes I){\rm E}]$,
with the trace taken over the input Hilbert space only, i.e.,
\begin{equation}
\mathcal{E}(\Omega_{in})=\sum_{i_{1},i_{2}=1}^{d_{in}}\sum_{i_{3},i_{4}=1}^{d_{out}}\chi_{i_{1}i_{2}i_{3}i_{4}}\left\langle \alpha_{i_{2}}\right|\Omega_{in}\left|\psi_{i_{1}}\right\rangle \left|\beta_{i_{3}}\right\rangle \left\langle \phi_{i_{4}}\right|.\label{sprocessdefinition}
\end{equation}
 When the four sets of basis states are chosen, the quantum process
$\mathcal{E}$ is completely defined by the coefficients $\chi_{i_{1}i_{2}i_{3}i_{4}}$.
Tomography of an unknown QP $\mathcal{E}$ is, therefore, to determine
the coefficients $\chi_{i_{1}i_{2}i_{3}i_{4}}$ with respect to the
chosen bases. The four sets of basis states could be chosen according
to our convenience.

This supplementary material is organized as follows. In Sec. A,
we present our main QPT scheme for the case when the couplings between
the system and the pointers are weak, and we derive the results in
the main text. In Sec. B, we show the advantages
of our scheme by comparing it with the existing QPT schemes. In Sec. C, we discuss the cases when the couplings are
not weak, we find that QPT is still possible and the problem can
actually be solved exactly when the observables we consider are all
projectors. In Sec D, we present an alternative QPT scheme for a single-qubit
gate, and show that the observables in the weak measurements need
not be projectors. In Sec. E, we extend our scheme to the case of
multi-particle processes and show that weak measurements of single-particle observables are sufficient for QPT of multi-particle processes.
In Sec. F, we present an ancilla-assisted version of our QPT scheme,
and it requires only one input state.

\subsection{Our main QPT scheme}

\begin{figure}
\includegraphics{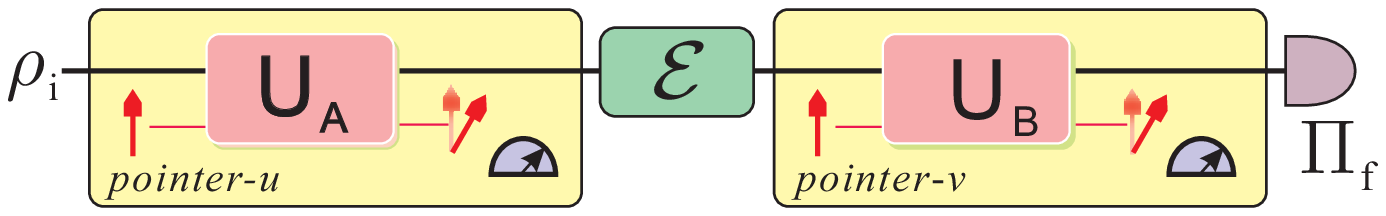}

\caption{Our main scheme of QPT}
\end{figure}

The key features of our scheme are depicted in Fig. 3. Firstly, we
choose a system of dimension $d_{in}$ and prepare it in a state $\rho_{i}$,
then weakly couple it with a pointer labeled by the letter {}``$u$''.
The interaction between them is described by the Hamiltonian $H_{A}=g\delta(t-t_{0})\hat{A}\otimes\hat{p}_{u}$,
which results in a unitary evolution $U_{A}=\exp(-ig\hat{A}\otimes\hat{p}_{u})$.
A small $g$ is chosen to ensure that the interaction is weak. Next,
the system undergoes the unknown QP $\mathcal{E}$. Then, the system
weakly interacts with a second pointer (labeled by {}``$v$'') and
the evolution is described by $U_{B}=\exp(-i\lambda\hat{B}\otimes\hat{p}_{v})$
where a small $\lambda$ is chosen to ensure a weakened interaction.
Finally, a projection $\Pi_{f}$ (the post-selection) is performed
on the system. We use the notation $(i\hat{A}\hat{B}f)$ to denote
this particular set of actions on the system.

For simplicity, suppose the pointers {}``$u$'' and {}``$v$''
are initially prepared in the states $\sigma_{u}$ and $\sigma_{v}$
respectively. For each pointer, we can read out either its $\hat{q}$
or $\hat{p}$. Therefore, there are four possible products of pointer
shifts, and we define their averages as: $r_{1}=\langle\hat{p}_{u}\otimes\hat{p}_{v}\rangle$,
$r_{2}=\langle\hat{p}_{u}\otimes\hat{q}_{v}\rangle$, $r_{3}=\langle\hat{q}_{u}\otimes\hat{p}_{v}\rangle$
and $r_{4}=\langle\hat{q}_{u}\otimes\hat{q}_{v}\rangle$ for each
set $(i\hat{A}\hat{B}f)$ of actions on the system. These $r$-values
can be directly obtained from the experiments.

The initial overall state of the combined system is $\varrho=\rho_{i}\otimes\sigma_{u}\otimes\sigma_{v}$,
and the final overall state is $\varrho'=\Pi_{f}U_{B}\mathcal{E}[U_{A}\varrho U_{A}^{\dagger}]U_{B}^{\dagger}\Pi_{f}$,
where $U_{A}\approx1-ig\hat{A}\otimes\hat{p}_{u}-\frac{1}{2}g^{2}\hat{A}^{2}\otimes\hat{p}_{u}^{2}$
and $U_{B}\approx1-i\lambda\hat{B}\otimes\hat{p}_{v}-\frac{1}{2}\lambda^{2}\hat{B}^{2}\otimes\hat{p}_{v}^{2}$.
Up to the second order of the coupling constants ($\lambda$ and $g$),
the final state is given as

\begin{equation}
\begin{aligned}\varrho'\approx\Pi_{f}\{ & \mathcal{E}(\rho_{i})\otimes\sigma_{u}\otimes\sigma_{v}\\
 & -ig[\mathcal{E}(\hat{A}\rho_{i})\otimes(\hat{p}_{u}\sigma_{u})-\mathcal{E}(\rho_{i}\hat{A})\otimes(\sigma_{u}\hat{p}_{u})]\otimes\sigma_{v}\\
 & -i\lambda[(\mbox{ }\hat{B}\mathcal{E}(\rho_{i})\mbox{ })\otimes(\hat{p}_{v}\sigma_{v})-(\mbox{ }\mathcal{E}(\rho_{i})\hat{B}\mbox{ })\otimes(\sigma_{v}\hat{p}_{v})]\otimes\sigma_{u}\\
 & -\frac{1}{2}g^{2}[\mathcal{E}(\hat{A}^{2}\rho_{i})\otimes(\hat{p}_{u}^{2}\sigma_{u})+\mathcal{E}(\rho_{i}\hat{A}^{2})\otimes(\sigma_{u}\hat{p}_{u}^{2})-2\mathcal{E}(\hat{A}\rho_{i}\hat{A})\otimes(\hat{p}_{u}\sigma_{u}\hat{p}_{u})]\otimes\sigma_{v}\\
 & -\frac{1}{2}\lambda^{2}[(\mbox{ }\hat{B}^{2}\mathcal{E}(\rho_{i})\mbox{ })\otimes(\hat{p}_{v}^{2}\sigma_{v})+(\mbox{ }\mathcal{E}(\rho_{i})\hat{B}^{2}\mbox{ })\otimes(\sigma_{v}\hat{p}_{v}^{2})-2(\mbox{ }\hat{B}\mathcal{E}(\rho_{i})\hat{B}\mbox{ })\otimes(\hat{p}_{v}\sigma_{v}\hat{p}_{v})]\otimes\sigma_{u}\\
 & -\lambda g[(\mbox{ }\hat{B}\mathcal{E}(\hat{A}\rho_{i})\mbox{ })\otimes(\hat{p}_{u}\sigma_{u})\otimes(\hat{p}_{v}\sigma_{v})-(\mbox{ }\mathcal{E}(\hat{A}\rho_{i})\hat{B}\mbox{ })\otimes(\hat{p}_{u}\sigma_{u})\otimes(\sigma_{v}\hat{p}_{v})\\
 & \mbox{ \ensuremath{}\ensuremath{}}-(\mbox{ }\hat{B}\mathcal{E}(\rho_{i}\hat{A})\mbox{ })\otimes(\sigma_{u}\hat{p}_{u})\otimes(\hat{p}_{v}\sigma_{v})+(\mbox{ }\mathcal{E}(\rho_{i}\hat{A})\hat{B}\mbox{ })\otimes(\sigma_{u}\hat{p}_{u})\otimes(\sigma_{v}\hat{p}_{v})]\mbox{ }\}\Pi_{f}
\end{aligned}
.\label{eq: state}
\end{equation}
 The trace of $\varrho'$ is the probability of obtaining $\Pi_{f}$
in the post-selection (when the input is $\rho_{i}$ and the two observables
of weak measurements are $\hat{A}$ and $\hat{B}$)
\begin{equation}
p_{f|i\hat{A}\hat{B}}=\mathrm{tr}(\varrho').\label{eq:paobability}
\end{equation}
 Then the normalized reduced density matrix of the two pointer can
be obtained by a partial trace over the system:
\[
\varrho_{uv}=\frac{1}{p_{f|i\hat{A}\hat{B}}}\mathrm{tr}_{s}(\varrho').
\]
 For convenience, we assume that the initial states of the two pointers
are the same and the average values of $\hat{q}$ and $\hat{p}$ of
each pointer are initially zero, i.e., $\sigma_{u}=\sigma_{v}=\sigma$
and $\mathrm{tr}(\hat{p}\sigma)=\mathrm{tr}(\hat{q}\sigma)=0$. It
can be seen that only the terms containing the factor $\lambda g$
in (\ref{eq: state}) contribute to the four $r$-values $r_{1}=\langle\hat{p}_{u}\otimes\hat{p}_{v}\rangle$,
$r_{2}=\langle\hat{p}_{u}\otimes\hat{q}_{v}\rangle$, $r_{3}=\langle\hat{q}_{u}\otimes\hat{p}_{v}\rangle$
and $r_{4}=\langle\hat{q}_{u}\otimes\hat{q}_{v}\rangle$. Here, the
expectation value, which is defined for each fixed set of actions
$(i\hat{A}\hat{B}f)$ on the system, is calculated with respect to
the final state $\rho_{uv}$ of the two pointers. For example, $r_{1}=\mathrm{tr}[(\hat{p}_{u}\otimes\hat{p}_{v})\varrho_{uv}]$,
where $\varrho_{uv}$ and $r_{1}$ both depend on the set of actions
$(i\hat{A}\hat{B}f)$ on the system. We drop the superscripts $(i\hat{A}\hat{B}f)$
when there is no confusion.

For convenience, we can define two $X$-values
\begin{equation}
\begin{aligned}X^{i\hat{A}\hat{B}f}=\mathrm{tr}[\Pi_{f}\hat{B}\mathcal{E}(\hat{A}\rho_{i})],\mbox{} & \tilde{X}^{i\hat{A}\hat{B}f}=\mathrm{tr}[\Pi_{f}\mathcal{E}(\hat{A}\rho_{i})\hat{B}]\end{aligned}
.\label{eq: define W}
\end{equation}
 The $X$-values can be viewed as a certain kind of joint weak values.
Then the terms in $\varrho_{uv}$ that contribute to the four $r$-values
can be expressed as

\begin{equation}
\begin{aligned}\varrho_{uv}\sim & -\frac{g\lambda}{p_{f|i\hat{A}\hat{B}}}[\mbox{ }X^{i\hat{A}\hat{B}f}(\hat{p}_{u}\sigma)\otimes(\hat{p}_{v}\sigma)-\tilde{X}^{i\hat{A}\hat{B}f}(\hat{p}_{u}\sigma)\otimes(\sigma\hat{p}_{v})\\
 & -(\tilde{X}^{i\hat{A}\hat{B}f})^{*}(\sigma\hat{p}_{u})\otimes(\hat{p}_{v}\sigma)+(\mbox{ }X^{i\hat{A}\hat{B}f})^{*}(\sigma\hat{p}_{u})\otimes(\sigma\hat{p}_{v})\mbox{ }]
\end{aligned}
.\label{eq:effective terms}
\end{equation}
 Let $c_{1}=\mathrm{tr}(\hat{q}\hat{p}\sigma)$ and $c_{2}=\mathrm{tr}(\hat{p}\hat{p}\sigma)$.
We drop the superscript ($i\hat{A}\hat{B}f$) when the context is
clear. The four $r$-values can be written as
\begin{equation}
\begin{pmatrix}\frac{r_{1}}{c_{2}^{2}}\\
\frac{r_{2}}{c_{2}}\\
\frac{r_{3}}{c_{2}}\\
r_{4}
\end{pmatrix}=\frac{-g\lambda}{p_{f|i\hat{A}\hat{B}}}\begin{pmatrix}1 & -1 & -1 & 1\\
c_{1} & -c_{1}^{*} & -c_{1} & c_{1}^{*}\\
c_{1} & -c_{1} & -c_{1}^{*} & c_{1}^{*}\\
c_{1}^{2} & -|c_{1}|^{2} & -|c_{1}|^{2} & c_{1}^{*2}
\end{pmatrix}\begin{pmatrix}X\\
\tilde{X}\\
\tilde{X}^{*}\\
X^{*}
\end{pmatrix}.\label{eq:inverse -1}
\end{equation}
Since the matrix in (\ref{eq:inverse -1}) is invertible, the $X$-values
can be expressed in terms of the $r$-values as
\begin{equation}
\begin{pmatrix}X\\
\tilde{X}\\
\tilde{X}^{*}\\
X^{*}
\end{pmatrix}=\frac{-p_{f|i\hat{A}\hat{B}}}{g\lambda(c_{1}-c_{1}^{*})^{2}}\begin{pmatrix}(\frac{c_{1}^{*}}{c_{2}})^{2} & -\frac{c_{1}^{*}}{c_{2}} & -\frac{c_{1}^{*}}{c_{2}} & 1\\
(\frac{|c_{1}|}{c_{2}})^{2} & -\frac{c_{1}^{*}}{c_{2}} & -\frac{c_{1}}{c_{2}} & 1\\
(\frac{|c_{1}|}{c_{2}})^{2} & -\frac{c_{1}}{c_{2}} & -\frac{c_{1}^{*}}{c_{2}} & 1\\
(\frac{c_{1}}{c_{2}})^{2} & -\frac{c_{1}}{c_{2}} & -\frac{c_{1}}{c_{2}} & 1
\end{pmatrix}\begin{pmatrix}r_{1}\\
r_{2}\\
r_{3}\\
r_{4}
\end{pmatrix} . \label{eq:inverse 2}
\end{equation}
Hence, we have proved Eq. (3) in the main text, which is just the first one of the above set of equations.

Now we focus on $X^{i\hat{A}\hat{B}f}$. From (\ref{sprocessdefinition}),
$X^{i\hat{A}\hat{B}f}$ can be expressed as

\begin{equation}
\begin{aligned}X^{i\hat{A}\hat{B}f} & =\mathrm{tr}[\Pi_{f}\hat{B}\mathcal{E}(\hat{A}\rho_{i})]\\
 & =\sum_{i'_{1},i'_{2}=1}^{d_{in}}\sum_{i'_{3},i'_{4}=1}^{d_{out}}\chi_{i'_{1}i'_{2}i'_{3}i'_{4}}\mathrm{tr}(\Pi_{f}\hat{B}|\beta_{i'_{3}}\rangle\langle\phi_{i'_{4}}|)\mbox{ }\langle\alpha_{i'_{2}}|\hat{A}\rho_{i}|\psi_{i'_{1}}\rangle
\end{aligned}
\label{eq:X expression}
\end{equation}
 When $\hat{A}=|\alpha_{i_{2}}\rangle\langle\alpha_{i_{2}}|$, $\hat{B}=|\beta_{i_{3}}\rangle\langle\beta_{i_{3}}|$,
$\rho_{i}=|\psi_{i_{1}}\rangle\langle\psi_{i_{1}}|$ and $\Pi_{f}=|\phi_{i_{4}}\rangle\langle\phi_{i_{4}}|$,
the index ($i\hat{A}\hat{B}f$) can be denoted as ($i_{1}i_{2}i_{3}i_{4}$).
Substituting the projectors into (\ref{eq:X expression}) we immediately
have
\begin{equation}
\begin{aligned}\chi_{i_{1}i_{2}i_{3}i_{4}} & =\frac{1}{\langle\phi_{i_{4}}|\beta_{i_{3}}\rangle\langle\alpha_{i_{2}}|\psi_{i_{1}}\rangle}X^{i_{1}i_{2}i_{3}i_{4}}\end{aligned}
.\label{eq:main 1}
\end{equation}
Thus, we have proved Eq. (4) in the main text.

The QP parameters are directly related to the experimental
data in an elegant way, as revealed in Eqs. (3) and (4) of the main text: the parameter $\chi_{i_{1}i_{2}i_{3}i_{4}}$ is
completely determined by the $X$-value $X^{i_{1}i_{2}i_{3}i_{4}}$, which, in
turn, is determined by five experimental values, i.e., the four $r$-values
and the probability $p_{f|i\hat{A}\hat{B}}$. This result is valid regardless
of the dimension of the system.

From the proof, we find that $\tilde{X}^{i\hat{A}\hat{B}f}$ can also
be used to determine the QP parameters. Let us rewrite the process
$\mathcal{E}$ as $\mathcal{E}(\Omega_{in})=\sum_{i_{1},i_{2}=1}^{d_{in}}\sum_{i_{3},i_{4}=1}^{d_{out}}\tilde{\chi}_{i_{1}i_{2}i_{3}i_{4}}\left\langle \alpha_{i_{2}}\right|\Omega_{in}\left|\psi_{i_{1}}\right\rangle \left|\phi_{i_{4}}\right\rangle \left\langle \beta_{i_{3}}\right|$,
where comparing with the representation in (\ref{sprocessdefinition}),
we have exchanged the roles of the two bases of $\mathcal{H}_{out}$.
Analogous calculation leads to the following result
\[
\tilde{\chi}_{i_{1}i_{2}i_{3}i_{4}}=\frac{1}{\langle\beta_{i_{3}}|\phi_{i_{4}}\rangle\langle\alpha_{i_{2}}|\psi_{i_{1}}\rangle}\tilde{X}^{i_{1}i_{2}i_{3}i_{4}}.
\]

\subsection{Our scheme versus the existing QPT schemes}

In this section, we compare the existing QPT schemes with ours. The
comparison will be under the assumption that $d_{in}=d_{out}=d$.
In the literature, some schemes cannot give all the QP parameters
\cite{qqscience symmetry}. And some schemes are designed for specific
physical systems, for example, the optical system \cite{qqscience optics}.
Here we focus on the four main schemes for complete QPT.

\paragraph*{Scheme a}

The standard QPT scheme \cite{qqQPT standard,qqstandard 2}. It requires
$d^{2}$ linearly independent input states, and the QP parameters
are determined in an indirect way via state tomography upon the output
states.

\paragraph*{Scheme b}

Ancilla-assisted process tomography \cite{qqAAPT,qqAAPT 2}. The idea
comes from the isomorphism between operators and quantum states, and
the scheme also relies on state tomography. To determine all QP parameters,
one needs only one full-rank mixed input state of the principal system
and the ancilla, similar to the ancilla-assisted version of our QPT scheme
presented in Sec. F of this Supplementary Information.

\paragraph*{Scheme c}

Direct characterization of QP \cite{qqDCQD,qqDCQD2}. This method is based
on the error correction theory. It requires an ancillary quantum system
and $d+2$ different entangled input states of the combined system,
with $d$ being a prime number. The measurements are performed jointly, in
entangled bases.

\paragraph*{Scheme d}

Selective and efficient QPT \cite{qqSEQPT}. Suppose the process $\mathcal{E}$
is written as $\mathcal{E}(\rho)=\sum_{a,b}\chi_{ab}E_{a}\rho E_{b}^{\dagger}$
where $\{ E_{a} | a=1\cdots d^{2} \}$ is a basis of operators. The success
of this scheme relies on a fact that $\frac{1}{K}\sum_{j}\langle\phi_{j}|\mathcal{E}(E_{a}^{\dagger}|\phi_{j}\rangle\langle\phi_{j}|E_{b})|\phi_{j}\rangle=\frac{d\chi_{ab}+\delta_{a,b}}{d+1}$,
where the summation is over all input states that form a
\textquotedblleft 2-design\textquotedblright, which has about $O(d^{2})$ elements. To
obtain $\chi_{ab}$, the input state $|\phi_{j}\rangle$ should be
acted upon by one of the four extra standard channels $(E_{a}\pm E_{b})^{\dagger}$
and $(E_{a}\pm iE_{b})^{\dagger}$ before undergoing the process $\mathcal{E}$,
and finally the output state should be projected onto the set of input
states $|\phi_{j}\rangle$.

For the physical resources, the numbers of different input states in
Scheme (a) and (d) are of order $d^{2}$, much larger than that of
ours. The input states required in Schemes (b) and (c) are much more difficult
to prepare. Additionally, Scheme (c) works only when $d$ is a prime;
otherwise, one should embed the system in a larger Hilbert space whose
dimension is a prime number.

As to parallelism, the numbers of setups required for Schemes (a)-(c)
are $O(d^{4})$, $O(d^{4})$ \cite{qqreview A} and $O(d^{2})$ \cite{qqDCQD2},
respectively. Scheme (d) is designed especially for partial QPT and
different strategies are needed to determine different QP parameters.
But for a complete QPT, Scheme (d) requires $O(d^{6})$ setups \cite{qqSEQPT}.
In contrast, our scheme requires only $d$ setups (for preparation
of the $d$ different initial states).

Then let us consider data processing and error accumulation. In Schemes
(a) and (b), the QP parameters are related to all the experimental
values via a $d^{4}\times d^{4}$ matrix \cite{qqNilsen,qqreview A}.
We use $\delta$ to roughly denote each systematical error produced
in the experimental operations, including the state preparation and
measurements. In Schemes (a), the systematic errors of an expectation
value arise from input-state preparation and the performance of the projective
measurements; in scheme (b) the errors of one expectation value arise
from the two measurements performed on the principal and the ancillary
systems. So the error of each estimated parameter $\chi$ in both schemes originates from $2d^{4}$
terms, i.e., $\delta(\chi)\sim 2d^{4}\delta$. In Scheme (c), the $d^{2}$
diagonal QP parameters can be directly obtained, but the $O(d^{4})$
non-diagonal parameters are related in a complex way to outcomes of
the joint measurements performed on $O(d)$ non-maximally entangled states
\cite{qqDCQD2}. Preparation of the entangled states and the collective
measurements on the composite system are more vulnerable to errors.
In Scheme (d), to give one QP parameter, $O(d^{2})$ different input
states and four extra standard QPs are required. Neglecting the possible
imperfections of extra QPs, the accumulated error can be expressed
as $\delta(\chi)\sim2d^{2}\delta$. In our QPT scheme, we have presented
in the main text that $\delta(\chi)\sim4\delta$. Therefore, our scheme
has an advantage over the other schemes with respect to the robustness
against error accumulation.

\subsection{A QPT scheme with exact solutions}

As weak measurements are the building blocks of our main scheme, the couplings between the pointers and the system should be weak.
However, for both theoretical and practical significance, it is interesting to discuss the cases when the couplings are not weak.

Actually, for the case of strong couplings, we still have a QPT scheme if $\hat{A}$ and $\hat{B}$ are both projective operators. In this case, we have $U_{A}=I_{su}+\hat{A}\otimes(e^{-ig\hat{p}_{u}}-I_{u})$ and $U_{B}=I_{sv}+\hat{B}\otimes(e^{-i\lambda\hat{p}_{v}}-I_{v})$.
Then, with the same setup as in Fig. 1, the final density matrix of the pointers is given exactly as
\begin{equation}
\begin{aligned}\mathrm{tr}_{s}(\varrho')= & \mathrm{tr}[\Pi_{f}\mathcal{E}(\rho_{i})]\sigma_{u}\otimes\sigma_{v}\\
\text{\ding{192}} & +\{\mbox{ }\mathrm{tr}[\Pi_{f}\mathcal{E}(\hat{A}\rho_{i})]\mbox{ }(e^{-ig\hat{p}_{u}}-I)\sigma_{u}+\mathrm{tr}[\Pi_{f}\mathcal{E}(\rho_{i}\hat{A})]\mbox{ }\sigma_{u}(e^{ig\hat{p}_{u}}-I)\mbox{ }\}\otimes\sigma_{v}\\
\text{\ding{192}} & +\{\mbox{ }\mathrm{tr}[\Pi_{f}\hat{B}\mathcal{E}(\rho_{i})]\mbox{ }(e^{-i\lambda\hat{p}_{v}}-I)\sigma_{v}+\mathrm{tr}[\Pi_{f}\mathcal{E}(\rho_{i})\hat{B}]\mbox{ }\sigma_{v}(e^{i\lambda\hat{p}_{v}}-I)\mbox{ }\}\otimes\sigma_{u}\\
\text{\ding{192}} & +\mathrm{tr}[\Pi_{f}\mathcal{E}(\hat{A}\rho\hat{A})]\mbox{ }(e^{-ig\hat{p}_{u}}-I)\sigma_{u}(e^{ig\hat{p}_{u}}-I)\otimes\sigma_{v}+\mathrm{tr}[\Pi_{f}\hat{B}\mathcal{E}(\rho_{i})\hat{B}]\mbox{ }(e^{-i\lambda\hat{p}_{v}}-I)\sigma_{v}(e^{i\lambda\hat{p}_{v}}-I)\otimes\sigma_{u}\\
\text{\ding{193}} & +\mathrm{tr}[\Pi_{f}\hat{B}\mathcal{E}(\hat{A}\rho_{i})]\mbox{ }(e^{-ig\hat{p}_{u}}-I)\sigma_{u}\otimes(e^{-i\lambda\hat{p}_{v}}-I)\sigma_{v}+\mathrm{tr}[\Pi_{f}\mathcal{E}(\hat{A}\rho_{i})\hat{B}]\mbox{ }(e^{-ig\hat{p}_{u}}-I)\sigma_{u}\otimes\sigma_{v}(e^{i\lambda\hat{p}_{v}}-I)\\
\text{\ding{193}} & +\mathrm{tr}[\Pi_{f}\hat{B}\mathcal{E}(\rho_{i}\hat{A})]\mbox{ }\sigma_{u}(e^{ig\hat{p}_{u}}-I)\otimes(e^{-i\lambda\hat{p}_{v}}-I)\sigma_{v}+\mathrm{tr}[\Pi_{f}\mathcal{E}(\rho_{i}\hat{A})\hat{B}]\mbox{ }\sigma_{u}(e^{ig\hat{p}_{u}}-I)\otimes\sigma_{v}(e^{i\lambda\hat{p}_{v}}-I)\\
\text{\ding{194}} & +\mathrm{tr}[\Pi_{f}\hat{B}\mathcal{E}(\hat{A}\rho_{i})\hat{B}]\mbox{ }(e^{-ig\hat{p}_{u}}-I)\sigma_{u}\otimes(e^{-i\lambda\hat{p}_{v}}-I)\sigma_{v}(e^{i\lambda\hat{p}_{v}}-I)\\
\text{\ding{194}} & +\mathrm{tr}[\Pi_{f}\hat{B}\mathcal{E}(\rho_{i}\hat{A})\hat{B}]\mbox{ }\sigma_{u}(e^{ig\hat{p}_{u}}-I)\otimes(e^{-i\lambda\hat{p}_{v}}-I)\sigma_{v}(e^{i\lambda\hat{p}_{v}}-I)\\
\text{\ding{195}} & +\mathrm{tr}[\Pi_{f}\hat{B}\mathcal{E}(\hat{A}\rho_{i}\hat{A})]\mbox{ }(e^{-ig\hat{p}_{u}}-I)\sigma_{u}(e^{ig\hat{p}_{u}}-I)\otimes(e^{-i\lambda\hat{p}_{v}}-I)\sigma_{v}\\
\text{\ding{195}} & +\mathrm{tr}[\Pi_{f}\mathcal{E}(\hat{A}\rho_{i}\hat{A})\hat{B}]\mbox{ }(e^{-ig\hat{p}_{u}}-I)\sigma_{u}(e^{ig\hat{p}_{u}}-I)\otimes\sigma_{v}(e^{i\lambda\hat{p}_{v}}-I)\\
\text{\ding{196}} & +\mathrm{tr}[\Pi_{f}\hat{B}\mathcal{E}(\hat{A}\rho_{i}\hat{A})\hat{B}]\mbox{ }(e^{-ig\hat{p}_{u}}-I)\sigma_{u}(e^{ig\hat{p}_{u}}-I)\otimes(e^{-i\lambda\hat{p}_{v}}-I)\sigma_{v}(e^{i\lambda\hat{p}_{v}}-I)
\end{aligned}
\label{eq:varrhofinal}
\end{equation}
 From the above expression, we know that the terms marked with \ding{192}
contribute only to the shift of one pointer, thus do not contribute
to the four $r$-values. The terms marked with \ding{193} \ding{194}
\ding{195} \ding{196} all contribute to the $r$-values. In
order to obtain the $X$-values in terms marked with
\ding{193} from the $r$-values, we should perform additional experiments
to determine the coefficients in the terms marked with \ding{194}
\ding{195} \ding{196}. In the following, let $\hat{A}=|\alpha_{i_{2}}\rangle\langle\alpha_{i_{2}}|$,
$\hat{B}=|\beta_{i_{3}}\rangle\langle\beta_{i_{3}}|$, $\rho_{i}=|\psi_{i_{1}}\rangle\langle\psi_{i_{1}}|$
and $\Pi_{f}=|\phi_{i_{4}}\rangle\langle\phi_{i_{4}}|$.

\paragraph{Determining the coefficient of \textmd{\textup{\ding{196}.}}}

Since $\hat{A}\rho_{i}\hat{A}=|\langle\alpha_{i_{2}}|\psi_{i_{1}}\rangle|^{2}\mbox{ }\hat{A}$,
$\hat{B}\Pi_{f}\hat{B}=|\langle\beta_{i_{3}}|\phi_{i_{4}}\rangle|^{2}\mbox{ }\hat{B}$,
the coefficient $\mathrm{tr}[\Pi_{f}\hat{B}\mathcal{E}(\hat{A}\rho_{i}\hat{A})\hat{B}]$
equals to $|\langle\alpha_{i_{2}}|\psi_{i_{1}}\rangle\langle\beta_{i_{3}}|\phi_{i_{4}}\rangle|^{2}\mbox{ }\mathrm{tr}[\hat{B}\mathcal{E}(\hat{A})]$.
The factor $|\langle\alpha_{i_{2}}|\psi_{i_{1}}\rangle\langle\beta_{i_{3}}|\phi_{i_{4}}\rangle|^{2}$
is fixed when the bases are chosen, we only need to determine the
factor $\mbox{ }\mathrm{tr}[\hat{B}\mathcal{E}(\hat{A})]$ experimentally.
For this purpose we need to perform the following experiment. First
we initialize the quantum system in the state $|\alpha_{i_{2}}\rangle\langle\alpha_{i_{2}}|=\hat{A}$,
then the system undergoes the QP $\mathcal{E}$ and its state turns
out to be $\mathcal{E}(\hat{A})$. At last, we project the system
onto $|\beta_{i_{3}}\rangle\langle\beta_{i_{3}}|=\hat{B}$ as a post-selection.
Then the value of $\mathrm{tr}[\hat{B}\mathcal{E}(\hat{A})]$ equals
to $p_{i_{3}|i_{2}}$, which is the probability of obtaining $|\beta_{i_{3}}\rangle$
in post-selection (when the input state is $|\alpha_{i_{2}}\rangle$).

\paragraph{Determining the coefficients of \textmd{\textup{\ding{194}.}}}

Similarly, we have $\mathrm{tr}[\Pi_{f}\hat{B}\mathcal{E}(\hat{A}\rho_{i})\hat{B}]=|\langle\beta_{i_{3}}|\phi_{i_{4}}\rangle|^{2}\mbox{ }\mathrm{tr}[\hat{B}\mathcal{E}(\hat{A}\rho_{i})]$
and $\mathrm{tr}[\Pi_{f}\hat{B}\mathcal{E}(\rho_{i}\hat{A})\hat{B}]=|\langle\beta_{i_{3}}|\phi_{i_{4}}\rangle|^{2}\mbox{ }\mathrm{tr}[\hat{B}\mathcal{E}(\rho_{i}\hat{A})]$.
To determine $\mathrm{tr}[\hat{B}\mathcal{E}(\hat{A}\rho_{i})]$ and
$\mathrm{tr}[\hat{B}\mathcal{E}(\rho_{i}\hat{A})]$, we need to perform
another experiment. First we prepare the system in an input state
$\rho_{i}=|\psi_{i_{1}}\rangle\langle\psi_{i_{1}}|$ and couple it
with a pointer labeled by {}``$u$'' via $U_{A}$, then the system
undergoes the process $\mathcal{E}$, which is followed by a projective
measurement onto $|\beta_{i_{3}}\rangle\langle\beta_{i_{3}}|=\hat{B}$.
The reduced density matrix of the pointers is finally given as
\[
\begin{aligned}\mathrm{tr}_{s}(\varrho'_{su})= & \mathrm{tr}[\hat{B}\mathcal{E}(\rho_{i})]\sigma_{u}+\\
 & \mathrm{tr}[\hat{B}\mathcal{E}(\hat{A}\rho_{i})]\mbox{ }(e^{-ig\hat{p}_{u}}-I)\sigma_{u}+\mathrm{tr}[\hat{B}\mathcal{E}(\rho_{i}\hat{A})]\mbox{ }\sigma_{u}(e^{ig\hat{p}_{u}}-I)+\\
 & \mathrm{tr}[\hat{B}\mathcal{E}(\hat{A}\rho_{i}\hat{A})]\mbox{ }(e^{-ig\hat{p}_{u}}-I)\sigma_{u}(e^{ig\hat{p}_{u}}-I)
\end{aligned}
\]
 The first term of $\mathrm{tr}_{s}(\varrho'_{su})$ does not contribute
to $\langle\hat{p}_{u}\rangle$ and $\langle\hat{q}_{u}\rangle$,
and the term $\mathrm{tr}[\hat{B}\mathcal{E}(\hat{A}\rho_{i}\hat{A})]$
has already been obtained when we try to determine the coefficient
of \ding{196} experimentally. So $\mathrm{tr}[\hat{B}\mathcal{E}(\hat{A}\rho_{i})]$
and $\mathrm{tr}[\hat{B}\mathcal{E}(\rho_{i}\hat{A})]$ can be easily
determined via the shifts of the pointer {}``$u$'', i.e., $\langle\hat{p}_{u}\rangle$
and $\langle\hat{q}_{u}\rangle$.

\paragraph{Determining the coefficients of \textmd{\textup{\ding{195}. }}}

The method to determine the coefficients of \ding{195} is similar
to that for \ding{194}, so we just present the experiment. First
we need to prepare a system in state $|\alpha_{i_{2}}\rangle\langle\alpha_{i_{2}}|=\hat{A}$
and the system undergoes the process $\mathcal{E}$, then we couple
it with a pointer via $U_{B}$ followed by a post-selection projecting
onto the state $|\phi_{i_{4}}\rangle$.

Having determined the coefficients in the terms maked with \ding{194}
\ding{195} \ding{196}, we can run our scheme in the main text
(see figure 1) to obtain the $r$-values, and then from the expression
of the final state of the pointers in (\ref{eq:varrhofinal}), we
can derive the $X$-values. From the $X$-values, we can obtain all
the QP parameters as in (\ref{eq:main 1}). Thus, we also have a complete
QPT scheme even when the couplings are not weak.

In conclusion, when the couplings between the system and the pointers are not weak, one needs three additional experiments to obtain each QP parameter. The directness of our scheme for this case is preserved, since the number of expectation values required to determine a QP parameter is still a constant, regardless of the dimension of the quantum system. However, there is a decline in parallelism. If the measurements of the observables $\{|\alpha_{i_{2}}\rangle\langle\alpha_{i_{2}}|\}_{i_{2}=1}^{d_{in}}$
and $\{|\beta_{i_{3}}\rangle\langle\beta_{i_{3}}|\}_{i_{3}=1}^{d_{out}}$ are all performed in a single setup, the exact solution of the reduced density matrix of any two pointers will be very complex, then we need more setups for additional experiments to obtain the X-values.

\subsection{Tomography of a single-qubit quantum process via weak measurements
of non-projective operators}

So far we have only considered weak measurements of projective operators.
Here, we show that projectors are not the only choice, and we present
a tomography scheme for a single-qubit quantum process via weak measurements
of non-projective operators.

Suppose that the single-qubit gate $\mathcal{E}$ is expressed in
the form of (\ref{sprocessdefinition}) where the four bases are all
chosen as the computational basis $\{|0\rangle,\mbox{ }|1\rangle\}$.
We prepare two input states $\rho_{0}=|0\rangle\langle0|$ and $\rho_{1}=|1\rangle\langle1|$,
choose both $\hat{A}$ and $\hat{B}$ to be $\sigma_{x}=|0\rangle\langle1|+|1\rangle\langle0|$,
and preform the post-selection onto two states: $\Pi_{0}=|0\rangle\langle0|$,
$\Pi_{1}=|1\rangle\langle1|$. Now, the notation $(i\hat{A}\hat{B}f)$
can be simplified as $(if)$ ($i,f\in\{0,1\}$). From the shifts of
pointers, we could also obtain the weak values $W_{if}^{\hat{A}}=\frac{1}{p_{f|i}}\mathrm{tr}\{\Pi_{f}\mathcal{E}[\hat{A}\rho_{i}]\}$
and $W_{if}^{\hat{B}}=\frac{1}{p_{f|i}}\mathrm{tr}\{\Pi_{f}\hat{B}\mathcal{E}(\rho_{i})]\}$
via
\[
\begin{aligned}W_{if}^{\hat{A}(\hat{B})}= & \frac{1}{2c_{2}\mathrm{Im}c_{1}g(\lambda)}(-c_{1}^{*}\langle\hat{p}_{u(v)}\rangle+c_{2}\langle\hat{q}_{u(v)}\rangle)\end{aligned}
\]
 Finally, the QP parameters are determined as $\chi_{iiff}=p_{f|i}$,
$\chi_{\bar{i}iff}=p_{f|i}W_{if}^{\hat{A}}$, $\chi_{ii\bar{f}f}=p_{f|i}W_{if}^{\hat{B}}$
and $\chi_{\bar{i}i\bar{f}f}=X^{if}$ ($\bar{i}$ stands for $1-i$,
and $\bar{f}$ stands for $1-f$). As in our main scheme, $X^{if}$
are obtained from the $r$-values, which are obtained directly from
the experiments.

Particularly, proper choices of $\hat{p}$, $\hat{q}$ and $\sigma$
will give $c_{1}$ a real part, and then the $X$-value can be obtained
with a single $r$-value: $r_{4}=\langle\hat{q}_{u}\otimes\hat{q}_{v}\rangle$.
When $\hat{A}=\hat{B}=\sigma_{x}$, the hermiticity of the quantum
states ensures that $X^{if}=\tilde{X}^{i\bar{f}}=(\tilde{X}^{\bar{i}f})^{*}=(X^{\bar{i}\bar{f}})^{*}$.
Suppose $c_{1}\equiv\mathrm{tr}(\hat{q}\hat{p}\sigma)$, and let $c_{1}^{2}=\alpha+i\beta$
with real numbers $\alpha,\mbox{ }\beta$. Then we have
\begin{equation}
\begin{aligned}p_{f|i}r_{4}^{fi} & =2(\alpha\mathrm{Re}X^{fi}-\beta\mathrm{Im}X^{fi})-2\sqrt{\alpha^{2}+\beta^{2}}\mathrm{Re}X^{\bar{f}i}\\
p_{\bar{f}|i}r_{4}^{\bar{f}i} & =2(\alpha\mathrm{Re}X^{\bar{f}i}-\beta\mathrm{Im}X^{\bar{f}i})-2\sqrt{\alpha^{2}+\beta^{2}}\mathrm{Re}X^{fi}\\
p_{f|\bar{i}}r_{4}^{f\bar{i}} & =2(\alpha\mathrm{Re}X^{\bar{f}i}+\beta\mathrm{Im}X^{\bar{f}i})-2\sqrt{\alpha^{2}+\beta^{2}}\mathrm{Re}X^{fi}\\
p_{\bar{f}|\bar{i}}r_{4}^{\bar{f}\bar{i}} & =2(\alpha\mathrm{Re}X^{fi}+\beta\mathrm{Im}X^{fi})-2\sqrt{\alpha^{2}+\beta^{2}}\mathrm{Re}X^{\bar{f}i}
\end{aligned}
\label{eq:r4 in one-qubit gate}
\end{equation}
 These equations can be written in a matrix form as:
\[
\frac{1}{2}\begin{pmatrix}p_{f|i}r_{4}^{fi}\\
p_{\bar{f}|i}r_{4}^{\bar{f}i}\\
p_{f|\bar{i}}r_{4}^{f\bar{i}}\\
p_{\bar{f}|\bar{i}}r_{4}^{\bar{f}\bar{i}}
\end{pmatrix}=\begin{pmatrix}\alpha & -\beta & -\sqrt{\alpha^{2}+\beta^{2}} & 0\\
-\sqrt{\alpha^{2}+\beta^{2}} & 0 & \alpha & -\beta\\
-\sqrt{\alpha^{2}+\beta^{2}} & 0 & \alpha & \beta\\
\alpha & \beta & -\sqrt{\alpha^{2}+\beta^{2}} & 0
\end{pmatrix}\begin{pmatrix}\mathrm{Re}X^{fi}\\
\mathrm{Im}X^{fi}\\
\mathrm{Re}X^{\bar{f}i}\\
\mathrm{Im}X^{\bar{f}i}
\end{pmatrix}.
\]
 The determinant of the matrix is $-4\beta^{4}$. Thus the $X$-values
can be obtained by the value of $r_{4}$ via an inverse matrix as
long as $\beta\neq0$. In order to have a nonzero imaginary part to
$c_{1}^{2}$, $c_{1}$ must have both non-zero real part and non-zero
imaginary part at the same time. Usually, $c_{1}\equiv\mathrm{tr}(\hat{q}\hat{p}\sigma)$
does have a nonzero imaginary part because $\hat{q}\hat{p}$ is not
a Hermite operator.

\subsection{Tomography of multi-particle QPs}

Application of our scheme for the tomography of multi-particle QPs
may involve multi-particle interaction, i.e., a pointer could be coupled
to all the particles. Here we show that our scheme for the tomography
of multi-particle processes can be accomplished with product input
states and weak measurements of single-particle observables only.
So each pointer needs to be coupled to only a single particle, and
the difficulty of the experiment is highly reduced.

Since we just need states in an orthonormal basis as the input, the
simplest choices of input states are obviously the product states
$\rho_{I}=\bigotimes_{j=1}^{N}\rho_{i}^{j}$ of the $N$ particles.
Here the $N$-particle process $\mathcal{E}$ can be expressed as
\begin{equation}
\mathcal{E}(\bigotimes_{j=1}^{N}\rho_{i}^{j})=\sum_{I_{1}I_{2}I_{3}I_{4}}\chi_{I_{1}I_{2}I_{3}I_{4}}\bigotimes_{j=1}^{N}\langle\alpha_{j,i_{2}}|\rho_{i}^{j}|\psi_{j,i_{1}}\rangle\mbox{ }|\beta_{j,i_{3}}\rangle\langle\phi_{j,i_{4}}|,\label{eq:many-body process}
\end{equation}
 where we use $j$ to label the different particles, use $(j,i_{t})$
to label the basis states of the Hilbert spaces of the $j$-particle,
the dummy index $I_{t}$ is short for $\{(1,i_{t}),\mbox{ }(2,i_{t})\cdots(N,i_{t})\}$
(we omit the range of summation), $\{|\psi_{j,i_{1}}\rangle\}$ is
an orthonormal basis of the input Hilbert space of the $j$-th particle,
and the other kets and bras have similar meanings.

As showed in Fig. 4, an $N$-particle QP has $N$ entries. In our
scheme, the input states of each entry compose an orthogonal basis
of the corresponding Hilbert space, and so do the post-selection states.
Let us mark them by $\rho_{i}^{j}\in\{|\psi_{j,i_{1}}\rangle\langle\psi_{j,i_{1}}|\}$
and $\Pi_{f}^{j}\in\{|\phi_{j,i_{4}}\rangle\langle\phi_{j,i_{4}}|\}$,
respectively. Before and after the system undergoing the QP $\mathcal{E}$,
we weakly couple each particle (the $j$-th particle) with pointers
whose initial states are $\sigma_{u_{j}}$ and $\sigma_{v_{j}}$,
respectively. The couplings are described via $U_{A_{j}}=\exp(-g_{j}\hat{A}_{j}\otimes\hat{p}_{u_{j}})$
and $U_{B_{j}}=\exp(-\lambda_{j}\hat{B}_{j}\otimes\hat{p}_{v_{j}})$.
For simplicity, we assume that the initial states of the pointers
are all $\sigma$ and $\mathrm{tr}(\hat{p}\sigma)=\mathrm{tr}(\hat{q}\sigma)=0$
in the following paragraphs.

\begin{figure}
\includegraphics[width=13cm]{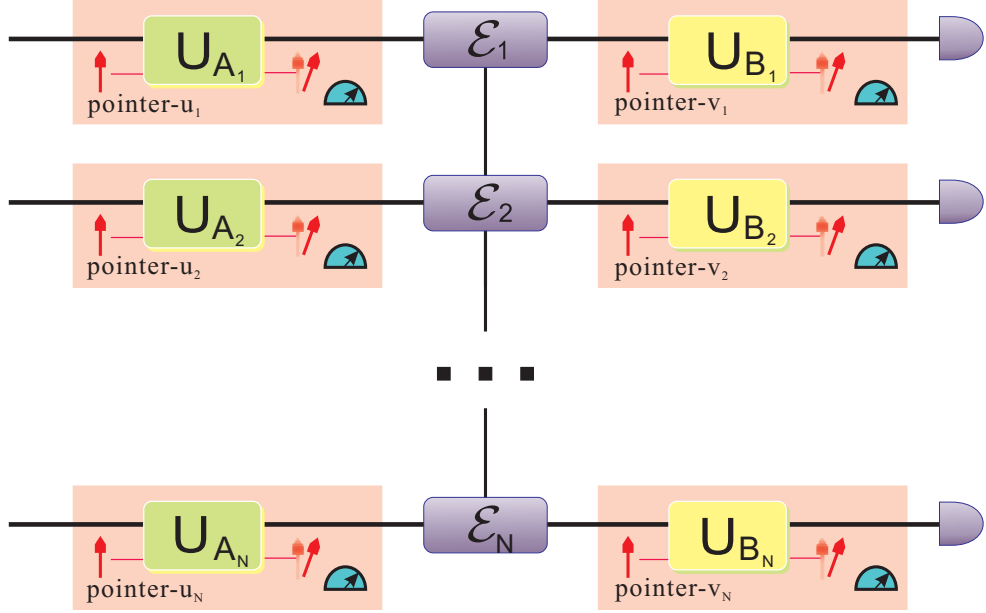}

\caption{An illustration of our QPT scheme for multi-particle quantum processes. }
\end{figure}

After the post-selection, the lowest-order terms of the reduced density
matrix of the pointers that contribute to the $4^{N}$ averages $\langle\otimes_{j=1}^{N}[\hat{p}_{u_{j}}(\hat{q}_{u_{j}})\otimes\hat{p}_{v_{j}}(\hat{q}_{v_{j}})]\rangle$
are

\begin{equation}
\begin{aligned}\frac{1}{p_{F|BAI}}\sum_{I_{1}I_{2}I_{3}I_{4}}\chi_{I_{1}I_{2}I_{3}I_{4}}\prod_{j=1}^{N}(-g_{j}\lambda_{j})\times\\
\{X_{j,i_{1}i_{2}i_{3}i_{4}}^{i\hat{A}\hat{B}f}(\hat{p}_{u_{j}}\sigma_{u_{j}})\otimes(\hat{p}_{v_{j}}\sigma_{v_{j}}) & -\tilde{X}_{j,i_{1}i_{2}i_{3}i_{4}}^{i\hat{A}\hat{B}f}(\hat{p}_{u_{j}}\sigma_{u_{j}})\otimes(\sigma_{v_{j}}\hat{p}_{v_{j}})\\
-(\tilde{X}_{j,i_{1}i_{2}i_{3}i_{4}}^{i\hat{A}\hat{B}f})^{*}(\sigma_{u_{j}}\hat{p}_{u_{j}})\otimes(\hat{p}_{v_{j}}\sigma_{v_{j}}) & +(X_{j,i_{1}i_{2}i_{3}i_{4}}^{i\hat{A}\hat{B}f})^{*}(\sigma_{u_{j}}\hat{p}_{u_{j}})\otimes(\sigma_{v_{j}}\hat{p}_{v_{j}})\}
\end{aligned}
,\label{eq:manybody uv state}
\end{equation}
 where $p_{F|BAI}$ is the probability of getting final states $\otimes_{j=1}^{N}\Pi_{f}^{j}$
in the post-selection, and
\begin{equation}
\begin{aligned}X_{j,i_{1}i_{2}i_{3}i_{4}}^{i\hat{A}\hat{B}f}= & \langle\alpha_{j,i_{2}}|\hat{A}_{j}\rho_{i}^{j}|\psi_{j,i_{1}}\rangle\mathrm{tr}[\Pi_{f}^{j}\hat{B}_{j}|\beta_{j,i_{3}}\rangle\langle\phi_{j,i_{4}}|],\\
\tilde{X}_{j,i_{1}i_{2}i_{3}i_{4}}^{i\hat{A}\hat{B}f}= & \langle\alpha_{j,i_{2}}|\hat{A}_{j}\rho_{i}^{j}|\psi_{j,i_{1}}\rangle\mathrm{tr}[\Pi_{f}^{j}|\beta_{j,i_{3}}\rangle\langle\phi_{j,i_{4}}|\hat{B}_{j}],
\end{aligned}
\label{eq:definition w multi}
\end{equation}

The $4^{N}$ expectation values $\langle\otimes_{j=1}^{N}[\hat{p}_{u_{j}}(\hat{q}_{u_{j}})\otimes\hat{p}_{v_{j}}(\hat{q}_{v_{j}})]\rangle$
are written as a column vector, by removing the factors $c_{2}^{2}$
and $(-1)^{N}\prod_{j=1}^{N}g_{j}\lambda_{j}$, one finds that this
column vector equals to
\begin{equation}
\frac{1}{p_{F|BAI}}\sum_{I_{1}I_{2}I_{3}I_{4}}\chi_{I_{1}I_{2}I_{3}I_{4}}\bigotimes_{j=1}^{N}\begin{pmatrix}1 & -1 & -1 & 1\\
c_{1} & -c_{1}^{*} & -c_{1} & c_{1}^{*}\\
c_{1} & -c_{1} & -c_{1}^{*} & c_{1}^{*}\\
c_{1}^{2} & -|c_{1}|^{2} & -|c_{1}|^{2} & c_{1}^{*2}
\end{pmatrix}\begin{pmatrix}X_{j,i_{1}i_{2}i_{3}i_{4}}^{i\hat{A}\hat{B}f}\\
\tilde{X}_{j,i_{1}i_{2}i_{3}i_{4}}^{i\hat{A}\hat{B}f}\\
(\tilde{X}_{j,i_{1}i_{2}i_{3}i_{4}}^{i\hat{A}\hat{B}f})^{*}\\
(X_{j,i_{1}i_{2}i_{3}i_{4}}^{i\hat{A}\hat{B}f})^{*}
\end{pmatrix}.\label{eq:inversion multi}
\end{equation}
 In (\ref{eq:inversion multi}), the constant matrix (invertible)
can be moved out of the summation, and then we are left with a vector
with $4^{N}$ elements
\[
\sum_{I_{1}I_{2}I_{3}I_{4}}\chi_{I_{1}I_{2}I_{3}I_{4}}\bigotimes_{j=1}^{N}(\mbox{ }X_{j,i_{1}i_{2}i_{3}i_{4}}^{i\hat{A}\hat{B}f},\mbox{ }\tilde{X}_{j,i_{1}i_{2}i_{3}i_{4}}^{i\hat{A}\hat{B}f},\mbox{ }(\tilde{X}_{j,i_{1}i_{2}i_{3}i_{4}}^{i\hat{A}\hat{B}f})^{*},\mbox{ }(X_{j,i_{1}i_{2}i_{3}i_{4}}^{i\hat{A}\hat{B}f})^{*})
\]
 which is related via an inversion matrix to the expectation values
$\langle\otimes_{j=1}^{N}[\hat{p}_{u_{j}}(\hat{q}_{u_{j}})\otimes\hat{p}_{v_{j}}(\hat{q}_{v_{j}})]\rangle$
that are directly obtained from the experiments. Particularly, the
first element of this vector is
\begin{equation}
X^{IABF}=\mathrm{tr}[\Pi_{F}B\mathcal{E}(A\rho_{I})]=\sum_{I'_{1}I'_{2}I'_{3}I'_{4}}\chi_{I'_{1}I'_{2}I'_{3}I'_{4}}\prod_{j=1}^{N}X_{j,i'_{1}i'_{2}i'_{3}i'_{4}}^{i\hat{A}\hat{B}f}.\label{eq: definition w1 multi}
\end{equation}
 Now we choose $\hat{A}_{j}=|\alpha_{j,i_{2}}\rangle\langle\alpha_{j,i_{2}}|$,
$\hat{B}_{j}=|\beta_{j,i_{3}}\rangle\langle\beta_{j,i_{3}}|$, then
the QP parameters can be obtained via
\begin{equation}
\chi_{I_{1}I_{2}I_{3}I_{4}}=\frac{X^{IABF}}{\prod_{j=1}^{N}\langle\alpha_{j,i_{2}}|\psi_{j,i_{1}}\rangle\langle\phi_{j,i_{4}}|\beta_{j,i_{3}}\rangle},\label{eq:key many-body}
\end{equation}
 so the QP can be expressed generally as
\[
\mathcal{E}(\rho_{N})=\sum_{I_{1}I_{2}I_{3}I_{4}}\chi_{I_{1}I_{2}I_{3}I_{4}}\mathrm{tr}[(\bigotimes_{j=1}^{N}|\psi_{j,i_{1}}\rangle\langle\alpha_{j,i_{2}}|)\mbox{ }\rho_{N}]\mbox{ }\bigotimes_{j=1}^{N}|\beta_{j,i_{3}}\rangle\langle\phi_{j,i_{4}}|.
\]

\subsection{Ancilla-assisted version of our QPT scheme}

In this section, we present an ancilla-assisted version of our main
QPT scheme, which requires only one input state.

We introduce an ancilla denoted by the letter {}``$r$''. In the
Schmidt bases, the input state we need can be written as:

\begin{equation}
\rho_{sr}=\sum_{i_{1},i_{2}=1}^{d_{in}}\gamma_{i_{1}i_{2}}|\alpha_{i_{2}}\rangle_{s}\langle\psi_{i_{1}}|\otimes|\alpha_{i_{2}}\rangle_{r}\langle\psi_{i_{1}}|,\label{eq:aast input}
\end{equation}

\begin{figure}
\includegraphics[width=13cm]{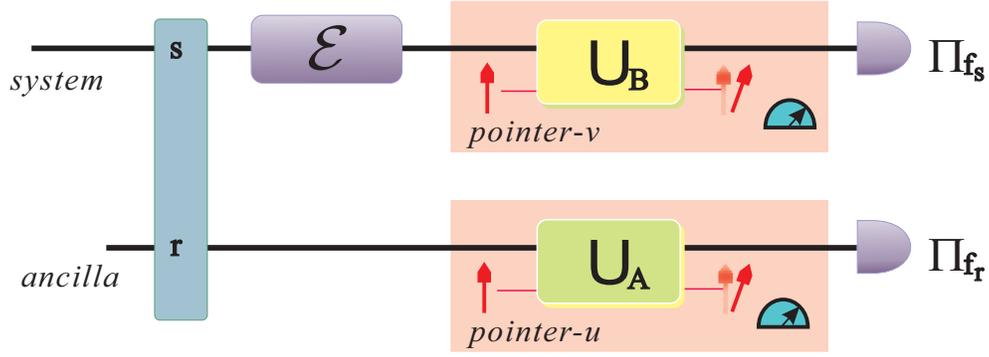}

\caption{An illustration of the ancilla-assisted version of our QPT scheme.}
\end{figure}

The ancilla-assisted version of our QPT scheme is illustrated in Fig.
3. The system undergoes the process, while we keep the ancilla invariant.
Then $\mathcal{E}$ translates the input state into

\[
\begin{aligned}\mathcal{E}_{s}\otimes I_{r}(\rho_{sr}) & =\sum_{i_{1},i_{2}=1}^{d_{in}}\sum_{i_{3},i_{4}=1}^{d_{out}}\chi_{i_{1}i_{2}i_{3}i_{4}}\gamma_{i_{1}i_{2}}|\beta_{i_{3}}\rangle_{s}\langle\phi_{i_{4}}|\otimes|\alpha_{i_{2}}\rangle_{r}\langle\psi_{i_{1}}|\end{aligned}
\]
 Then we couple the system with pointer {}``$v$'' weakly via the
evolution operator $U_{B}$, and couple the ancilla with pointer {}``$u$''
weakly via $U_{A}$, followed by post-selections $\Pi_{f_{s}}$ and
$\Pi_{f_{r}}$ on the system and the ancilla, respectively. So the
final overall state of the system, the ancilla and the pointers is

\[
\varrho_{sruv}=\Pi_{f_{s}}\Pi_{f_{r}}U_{B}U_{A}\{[\mathcal{E}_{s}\otimes I_{r}(\rho_{sr})]\otimes\sigma_{u}\otimes\sigma_{v}\}U_{A}^{\dagger}U_{B}^{\dagger}\Pi_{f_{r}}\Pi_{f_{s}},
\]
 The probability of getting the post-selected state $\Pi_{f_{s}}\otimes\Pi_{f_{r}}$
is
\[
p_{f_{s}f_{r}|\hat{A}\hat{B}}=\mathrm{tr}(\varrho_{sruv})
\]
 To the second order, the reduced density matrix of the two pointers
is

\begin{equation}
\begin{aligned}\varrho_{uv} & =\frac{1}{p_{f_{s}f_{r}|\hat{A}\hat{B}}}\mathrm{tr}_{sr}(\varrho_{sruv})\\
 & =\frac{1}{p_{f_{s}f_{r}|\hat{A}\hat{B}}}\sum_{i_{1}i_{2}i_{3}i_{4}}\chi_{i_{1}i_{2}i_{3}i_{4}}\gamma_{i_{1}i_{2}}\{\mathrm{tr}(\Pi_{f_{r}}|\alpha_{i_{2}}\rangle_{r}\langle\psi_{i_{1}}|)\mathrm{tr}(\Pi_{f_{s}}|\beta_{i_{3}}\rangle_{s}\langle\phi_{i_{4}}|)\sigma_{u}\otimes\sigma_{v}\\
 & -ig\mathrm{tr}(\Pi_{f_{s}}|\beta_{i_{3}}\rangle_{s}\langle\phi_{i_{4}}|)\mbox{ }[\mathrm{tr}(\Pi_{f_{r}}\hat{A}|\alpha_{i_{2}}\rangle_{r}\langle\psi_{i_{1}}|)\mbox{ }\hat{p}_{u}\sigma_{u}+\mathrm{tr}(\Pi_{f_{r}}|\alpha_{i_{2}}\rangle_{r}\langle\psi_{i_{1}}|\hat{A})\mbox{ }\sigma_{u}\hat{p}_{u}]\otimes\sigma_{v}\\
 & -i\lambda\mathrm{tr}(\Pi_{f_{r}}|\alpha_{i_{2}}\rangle_{r}\langle\psi_{i_{1}}|)\mbox{ }[\mathrm{tr}(\Pi_{f_{s}}\hat{B}|\beta_{i_{3}}\rangle_{s}\langle\phi_{i_{4}}|)\mbox{ }\hat{p}_{v}\sigma_{v}-\mathrm{tr}(\Pi_{f_{s}}|\beta_{i_{3}}\rangle_{s}\langle\phi_{i_{4}}|\hat{B})\mbox{ }\sigma_{v}\hat{p}_{v}]\otimes\sigma_{u}\\
 & -\frac{1}{2}g^{2}\mathrm{tr}(\Pi_{f_{s}}|\beta_{i_{3}}\rangle_{s}\langle\phi_{i_{4}}|)\mbox{ }[\mathrm{tr}(\Pi_{f_{r}}\hat{A}^{2}|\alpha_{i_{2}}\rangle_{r}\langle\psi_{i_{1}}|)\mbox{ }\hat{p}_{u}^{2}\sigma_{u}+\mathrm{tr}(\Pi_{f_{r}}|\alpha_{i_{2}}\rangle_{r}\langle\psi_{i_{1}}|\hat{A}^{2})\mbox{ }\sigma_{u}\hat{p}_{u}^{2}\\
 & \mbox{\ensuremath{}}-2\mathrm{tr}(\Pi_{f_{r}}\hat{A}|\alpha_{i_{2}}\rangle_{r}\langle\psi_{i_{1}}|\hat{A})\mbox{ }\hat{p}_{u}\sigma_{u}\hat{p}_{u}]\otimes\sigma_{v}\\
 & -\frac{1}{2}\lambda^{2}\mathrm{tr}(\Pi_{f_{r}}|\alpha_{i_{2}}\rangle_{r}\langle\psi_{i_{1}}|)\mbox{ }[\mathrm{tr}(\Pi_{f_{s}}\hat{B}^{2}|\beta_{i_{3}}\rangle_{s}\langle\phi_{i_{4}}|)\mbox{ }\hat{p}_{v}^{2}\sigma_{v}+\mathrm{tr}(\Pi_{f_{s}}|\beta_{i_{3}}\rangle_{s}\langle\phi_{i_{4}}|\hat{B}^{2})\mbox{ }\sigma_{v}\hat{p}_{v}^{2}\\
 & \mbox{\ensuremath{}}-2\mathrm{tr}(\Pi_{f_{s}}\hat{B}|\beta_{i_{3}}\rangle_{s}\langle\phi_{i_{4}}|\hat{B})\mbox{ }\hat{p}_{v}\sigma_{v}\hat{p}_{v})]\otimes\sigma_{u}\\
 & -g\lambda[\mathrm{tr}(\Pi_{f_{r}}\hat{A}|\alpha_{i_{2}}\rangle_{r}\langle\psi_{i_{1}}|)\mbox{ }\mathrm{tr}(\Pi_{f_{s}}\hat{B}|\beta_{i_{3}}\rangle_{s}\langle\phi_{i_{4}}|)\mbox{ }\hat{p}_{u}\sigma_{u}\otimes\hat{p}_{v}\sigma_{v}-\mathrm{tr}(\Pi_{f_{r}}\hat{A}|\alpha_{i_{2}}\rangle_{r}\langle\psi_{i_{1}}|)\mbox{ }\mathrm{tr}(\Pi_{f_{s}}|\beta_{i_{3}}\rangle_{s}\langle\phi_{i_{4}}|\hat{B})\mbox{ }\hat{p}_{u}\sigma_{u}\otimes\sigma_{v}\hat{p}_{v}\\
 & -\mathrm{tr}(\Pi_{f_{r}}|\alpha_{i_{2}}\rangle_{r}\langle\psi_{i_{1}}|\hat{A})\mbox{ }\mathrm{tr}(\Pi_{f_{s}}\hat{B}|\beta_{i_{3}}\rangle_{s}\langle\phi_{i_{4}}|)\mbox{ }\sigma_{u}\hat{p}_{u}\otimes\hat{p}_{v}\sigma_{v}+\mathrm{tr}(\Pi_{f_{r}}|\alpha_{i_{2}}\rangle_{r}\langle\psi_{i_{1}}|\hat{A})\mbox{ }\mathrm{tr}(\Pi_{f_{s}}|\beta_{i_{3}}\rangle_{s}\langle\phi_{i_{4}}|\hat{B})\mbox{ }\sigma_{u}\hat{p}_{u}\otimes\sigma_{v}\hat{p}_{v}]\}
\end{aligned}
\label{eq:1}
\end{equation}
 Only the terms with factor $g\lambda$ contribute to the four expectation
values of the pointers $\langle\hat{p}_{u}\otimes\hat{p}_{v}\rangle$,
$\langle\hat{p}_{u}\otimes\hat{q}_{v}\rangle$, $\langle\hat{q}_{u}\otimes\hat{p}_{v}\rangle$
and $\langle\hat{q}_{u}\otimes\hat{q}_{v}\rangle$. Define two $X$-values
as:
\begin{equation}
\begin{aligned}X^{f_{r}\hat{A}\hat{B}f_{s}}= & \sum_{i_{1}i_{2}i_{3}i_{4}}\chi_{i_{1}i_{2}i_{3}i_{4}}\gamma_{i_{1}i_{2}}\mathrm{tr}(\Pi_{f_{r}}\hat{A}|\alpha_{i_{2}}\rangle_{r}\langle\psi_{i_{1}}|)\mathrm{tr}(\Pi_{f_{s}}\hat{B}|\beta_{i_{3}}\rangle_{s}\langle\phi_{i_{4}}|),\\
\tilde{X}^{f_{r}\hat{A}\hat{B}f_{s}}= & \sum_{i_{1}i_{2}i_{3}i_{4}}\chi_{i_{1}i_{2}i_{3}i_{4}}\gamma_{i_{1}i_{2}}\mathrm{tr}(\Pi_{f_{r}}\hat{A}|\alpha_{i_{2}}\rangle_{r}\langle\psi_{i_{1}}|)\mathrm{tr}(\Pi_{f_{s}}|\beta_{i_{3}}\rangle_{s}\langle\phi_{i_{4}}|\hat{B}),
\end{aligned}
\label{eq: assisted W definition}
\end{equation}
 They can be obtained from the four expectation values via the inversion
relation given in (\ref{eq:inverse 2}).

Suppose the post-selections of the system and the ancilla are projections
onto $\Pi_{f_{s}}\in\{|\phi_{i_{4}}\rangle\langle\phi_{i_{4}}|\}_{i_{4}=1}^{d_{out}}$
and $\Pi_{f_{r}}\in\{|\psi_{i_{1}}\rangle\langle\psi_{i_{1}}|\}_{i_{1}=1}^{d_{in}}$,
respectively. When $\hat{A}=|\alpha_{i_{2}}\rangle\langle\alpha_{i_{2}}|$,
$\hat{B}=|\beta_{i_{3}}\rangle\langle\beta_{i_{3}}|$, $\Pi_{f_{r}}=|\psi_{i_{1}}\rangle\langle\psi_{i_{1}}|$
and $\Pi_{f_{s}}=|\phi_{i_{4}}\rangle\langle\phi_{i_{4}}|$, we denote
the set of indexes $(f_{r}\hat{A}\hat{B}f_{s})$ as $(i_{1}i_{2}i_{3}i_{4})$.
We now focus on $X^{i_{1}i_{2}i_{3}i_{4}}$ for a fixed set of indexes
$(i_{1}i_{2}i_{3}i_{4})$. From (\ref{eq: assisted W definition}),
it is apparent to see that

\begin{equation}
\chi_{i_{1}i_{2}i_{3}i_{4}}=\frac{X^{i_{1}i_{2}i_{3}i_{4}}}{\gamma_{i_{1}i_{2}}\langle\phi_{i_{4}}|\beta_{i_{3}}\rangle\langle\psi_{i_{1}}|\alpha_{i_{1}}\rangle}.\label{eq:main assisted}
\end{equation}
 Therefore, each QP parameter $\chi_{i_{1}i_{2}i_{3}i_{4}}$ is directly
related to the $X$-value $X^{i_{1}i_{2}i_{3}i_{4}}$, which, in turn,
is related to the experimental expectation values via (\ref{eq:inverse 2}).
$\tilde{X}$ can also be used to determine the QP parameters in a
similar way.

If the QP is a multi-particle one, the ancilla-assisted version can
also be accomplished with weak measurements of single-particle observables
only. The derivation is similar to that in Section E, and we don't
repeat it here. In brief, the strategy showed in Fig. 4 contains our
main scheme as basic elements, we just need to replace each of them
with the ancilla-assisted version showed in Fig. 5. And then the input
state we need is just a product of $N$ bipartite states.

\end{document}